\documentclass{article}

\usepackage{PRIMEarxiv}

\usepackage[utf8]{inputenc} 
\usepackage[T1]{fontenc}    
\usepackage{hyperref}       
\usepackage{url}            
\usepackage{booktabs}       
\usepackage{amsmath, amssymb}
\usepackage{nicefrac}       
\usepackage{microtype}      
\usepackage{lipsum}
\usepackage{fancyhdr}       
\usepackage{graphicx}       
\graphicspath{{media/}}     
\usepackage{natbib}

\title{Confidence intervals for the Cox model test error from cross-validation}

\author{
  Min Woo Sun \\
  Department of Biomedical Data Science \\
  Stanford University \\
  \texttt{minwoos@stanford.edu} \\
   \And
  Robert Tibshirani \\
  Department of Biomedical Data Science \\
  Department of Statistics \\
  Stanford University \\
  \texttt{tibs@stanford.edu} \\
}

\begin{document}
\maketitle

\begin{abstract}
Cross-validation (CV) is one of the most widely used techniques in statistical learning for estimating the test error of a model, but its behavior is not yet fully understood. It has been shown that  standard confidence intervals for test error using estimates from CV may  have coverage below nominal levels. This phenomenon occurs because each sample is used in both the training and testing procedures during CV and as a result, the CV estimates of the errors become correlated. Without accounting for this correlation, the estimate of the variance is smaller than it should be. One way to mitigate this issue is by estimating the mean squared error of the prediction error instead using nested CV. This approach has been shown to achieve superior coverage compared to intervals derived from standard CV. In this work, we generalize the nested CV idea to the Cox proportional hazards model and explore various choices of test error for this setting.
\end{abstract}


\section{Introduction}\label{sec1}
Cox models are deployed ubiquitously for survival analysis, for example for assessing the success of chemohormonal therapy in prostate cancer, or estimating customer churn \citep{chemo, churn}. Cox proportional hazards (PH) model is a class of survival models often used to model the relationship between time to event and a set of covariates. Survival data is typically characterized by the following properties. The response variable $Y$ is time to occurrence of an event, often time to failure or death. Observations can be censored--denoted with an indicator variable ($\delta = 0$ if censored, $\delta = 1$ if the event occurs). For instance, if a participant in a clinical trial drops out during the study resulting in an unknown survival time, then the observation is right censored. Lastly, there are predictor variables of interest $X = (x_{1}, x_{2}, \cdots, x_{k})^{T}$. The Cox PH regression model aims to understand the relationship between the response and a given set of predictors. Due to censoring, standard approaches like ordinary least squares regression cannot be employed in this setting. The proportional hazards model was introduced by Cox for solving this particular problem \citep{Cox_1972}. The Cox PH model is defined as follows,

\begin{eqnarray}
\lambda_{i}(t) = \lambda_{0}(t) e^{x_{i}^{T}\beta} 
\end{eqnarray}
where $\lambda_{i}(t)$ represents the hazards for individual $i$ at time $t$ and $\lambda_{0}(t)$ is the baseline hazard (i.e. the hazard when $X=(0, 0, \cdots, 0)^{T}$) that is shared across all the individuals in the data. $x_{i}$ is the vector of predictors measured on individual $i$, and $\beta=(\beta_{1}, \beta_{2}, \cdots, \beta_{k})$ the corresponding coefficients. The exponential function is a common choice for defining the relative risk, which can be thought of as the proportionate change in risk based on the predictors. 
One approach to fitting the Cox model is by maximizing the partial likelihood (PL) \citep{Cox_1975}. For simplicity, we can assume distinct event times to avoid ties. The PL is defined as, 

\begin{eqnarray}
L(\beta) = \prod_{j=1}^{m} \frac{e^{x_{j}^{T} \beta }}{\sum_{i \in R(j)} e^{x_{i}^{T} \beta }}
\end{eqnarray}
where $R(j)$ is the risk set representing individuals that are at risk of the event at the time of the event occurring for individual $j$. Note that the baseline hazard $\lambda_0(t)$ gets canceled out in the PL formulation. As such, the baseline hazard does not need to be specified for inference, making the approach semi-parametric and the problem becomes independent of time. Furthermore, regularization techniques like $l_{1}$ lasso and $l_{2}$ ridge penalties can be incorporated into the PL. The corresponding solution for the coefficients can be achieved through optimization procedures like coordinate descent, which has been incorporated in R package {\bf glmnet} \citep{Simon_2011}. 

It is important to accurately quantify uncertainty and assess the Cox model’s performance especially for its prevalent use in the biomedical field. Throughout this work, we refer to model performance objectives like the concordance index and the log partial likelihood as prediction error or test error. A common approach for fitting and evaluating statistical learning models involves cross validation (CV) \citep{Stone_1974}. $K$-fold CV is a simple technique that randomly splits data into $K$ equally sized folds then trains a model on $K-1$ folds and tests on the held-out fold rather than relying on a single train-test split. By repeating this procedure for all $K$ folds, CV will generate $K$ prediction error estimates of the model. As such, it is possible to estimate the variance based on these CV-derived estimates and construct a confidence interval around the point estimate.

\begin{figure}
\centerline{\includegraphics[width=350pt,height=23pc]{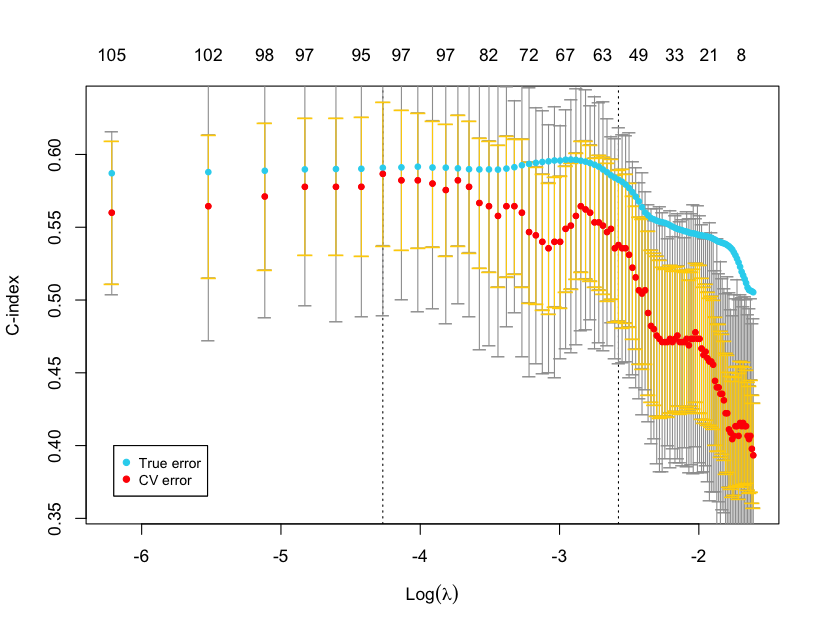}}
\caption{ {\it This figure demonstrates that naive confidence intervals (yellow) do not have adequate coverage and must be adjusted to reach nominal coverage. We run CV and nested CV on a train set of size $n=100$ observations and $p=150$ features using a $\ell_1$-regularized Cox PH model to construct 90\% confidence intervals. The solution path is plotted as a function of $\log(\lambda)$, where $\lambda$ is weight on the penalty term. The number of selected features is shown along the top of the plot. We estimated the out-of-sample C-index using a a test set of size n=1000. The models corresponding to the highest C-index and one SE rule are indicated by the vertical broken lines.} }
\end{figure}

Confidence intervals constructed from CV estimates of the prediction error are often much smaller than their purported nominal coverage as shown in Figure 1 \citep{Bates_2021}. This phenomenon occurs due to the correlation induced between the prediction errors by repeated use of samples for both training and testing in the CV procedure. More precisely, the estimate of the variance used to compute the standard error assumes that the observed prediction errors are independent, thus overlooking the covariance terms. As a result, the estimate of the variance ends up being smaller than they should be for building confidence intervals with proper coverage. The problem of narrow confidence intervals is exacerbated when there are more predictors than samples (p > n), which is a common setting for biological data such as gene expression data.

In order to address this problem, Bates {\it et al.} (2021) recently proposed using the estimate of the mean squared error (MSE) of the prediction error, rather than the sample variance of the prediction error \citep{Bates_2021}. MSE is an appropriate choice as it can be decomposed into variance and bias terms, where bias is usually small, thus making MSE a conservative estimate for the squared standard error \citep{Efron_1997}. Furthermore, Bates {\it et al.} (2021) introduces a convenient lemma that allows the MSE to be directly estimated from the data through nested CV without making any asymptotic or modeling assumptions as we have seen in prior related works constructing asymptotically-exact confidence intervals \citep{Bayle_2020}. We delve deeper into the lemma and the nested CV procedure in Section 3. Simulation experiments and real data examples from Bates {\it et al.} (2021) demonstrate that the confidence intervals constructed with MSE estimate achieves valid coverage for linear and logistic regression models in both low-dimensional and high-dimensional settings.

In this paper we demonstrate the efficacy of nested CV for constructing proper confidence intervals for test error in the survival analysis setting. We explore various choices of test error for assessing the model performance such as the concordance index and the log PL. We present a technique using nested CV to construct valid confidence intervals for the concordance index (C-index). While this work focuses on the Cox PH model, this nested CV approach can be implemented for other survival models.

\section{Measuring model performance}\label{sec2}
\subsection{Concordance Index (C-index)}
Evaluating a predictive model in the presence of censored data is not simple because measures like MSE cannot be computed. We instead use Harrell’s concordance index or C-index based on Kendall--Goodman-Kruskal-Somers type rank correlation as introduced by Harrell and implemented in {\bf glmnet} \citep{HARRELL_1996, glmnet}. The C-index measures a model’s predictive discrimination power or more precisely, the proportion of observation pairs for which the corresponding outcomes and predictions are concordant, i.e. for a pair $(i, j)_{i \neq j}$, $x_{i} > x_{j}, f(x_{i}) > f(x_{j})$ or $x_{i} < x_{j}, f(x_{i}) < f(x_{j})$, where $x$ is an observation and $f(x)$ is the prediction from a model. In the context of the Cox model, given a pair of subjects $i$ and $j$, a concordant pair would mean that $i$’s predicted risk is greater than $j$’s when the survival time is less than $j$’s, or vice versa. We exclude pairs that cannot be ordered--pairs for which both samples had their event occur simultaneously or if for one of the samples the event occurs but the event status for the other sample is unknown due to right censoring (e.g. patient dropped out of the study) prior to the first sample's event. The C-index ranges between $0$ and $1$ where a value of $1$ indicates perfect separation and a value of $0.5$ means no predictive discrimination, i.e. randomly guessing for each sample. 

The C-index presents a unique challenge for computing the variance of the measure because the C-index is defined for the set of observations not for an individual sample. The variance of the C-index is necessary for estimating the variance of the held-out error in the nested CV algorithm. In the case of MSE, one can simply compute the sample variance of the CV errors. The variance of the C-index can be estimated using the Infinitesimal Jackknife method as implemented in the “survival” package. A detailed description of the approach can be found in the package vignette: \url{ https://cran.r-project.org/web/packages/survival/vignettes/concordance.pdf}

\subsection{Partial Likelihood (PL)}
\begin{figure}
\centerline{\includegraphics[width=370pt,height=20pc]{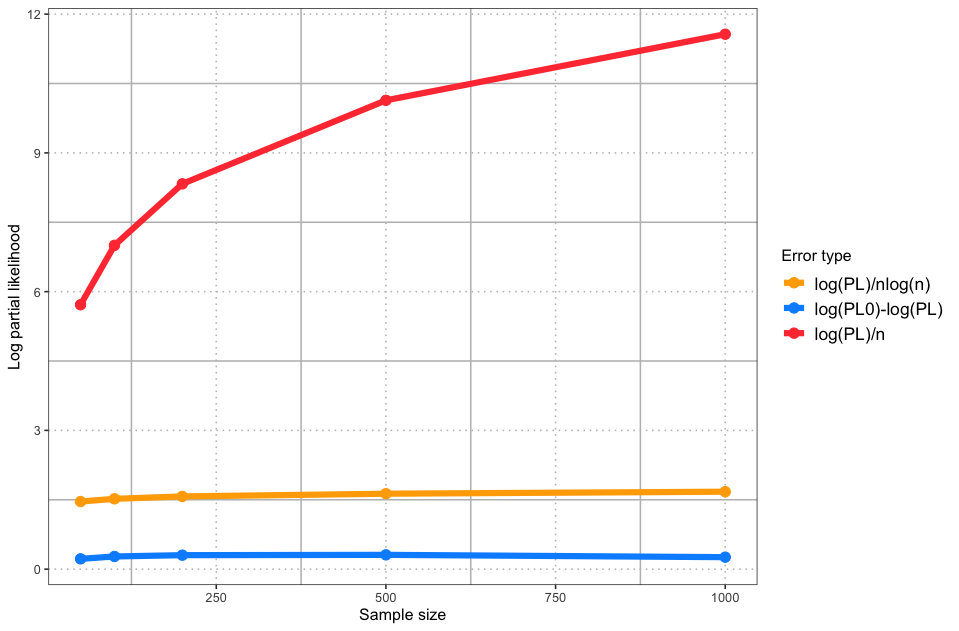}}
\caption{{\it This plot shows the relationship between sample size and various test errors for the Cox model. It is clear that $\log(PL) / n$ increases with sample size despite scaling by $n^{-1}$ while the other measures remain constant. As a result, $\log(PL) / n$ is not a viable candidate for test error.}}
\end{figure}

An alternative choice for prediction error for the Cox model is the log partial likelihood, which is also used as the loss function for fitting the Cox PH model as described in Section 2,
\begin{eqnarray}
\ell(\beta) = \sum_{j=1}^{m} x_{j}^{T} \beta - \log({\sum_{i \in R(j)} e^{x_{i}^{T} \beta}})
\end{eqnarray}
Since the PL is a product, as the number of samples increases its magnitude grows; therefore, its value needs to be scaled by the sample size. However, simply dividing the log PL by the sample size $n$ still exhibits an increasing behavior because $\log(PL)/n$ does not converge to a finite expectation. When the sample size $n$ is big, approximately $\mathbb{E}[\ell(\beta)/n] = c_{1} + c_{2} \cdot \log(n)$. This is an issue for evaluating confidence interval coverage as the the test error will increase with larger held out samples. As a result, $\ell(\beta)/n$ is not a viable measure to use to evaluate model performance. In order to avoid this problem, one can  instead use
\begin{eqnarray}
\frac{\ell(\beta)}{n\log(n)}
\end{eqnarray}
 or
 \begin{eqnarray}
\ell(\beta_{\text{null}}) - \ell(\beta)
\end{eqnarray} 
where $\ell(\beta_{null})$ is the log PL from the Cox model trained on data with no signal (i.e. covariates are zero). In Figure 2 while $\log(PL)/n$ increases with sample size, the latter two do not.  However we instead focus on C-index for test set evaluation here, because it is more interpretable. As is standard, we do employ PL for cross-validation (next section). 

\subsection{Cross-validating the Partial Likelihood}
Using PL poses a challenge for traditional CV as the PL is defined for a set of observations, not just a single sample. Furthermore, due to censoring in survival analysis, a left out sample in a leave-one-out CV procedure (fold size $k = n$) can be ill-defined \citep{Simon_2011}. In this work we focus on the nested CV implementation for the C-index, but CV and nested CV procedures for the log PL can be implemented following the leave-one-out CV procedure introduced by Houwelingen {\it et al.} (2006) \citep{Houwelingen}. The CV log PL is defined as,

\begin{eqnarray}
\widehat{\text{CV}}_{\text{PL}}(\lambda) = \sum_{i=1}^{n} \ell(\hat{\beta}_{\lambda}^{(-i)}) - \ell_{-i}(\hat{\beta}_{\lambda}^{(-i)})
\end{eqnarray}
where $\hat{\beta}_{\lambda}^{(-i)}$ refers to the optimal fit excluding sample $i$ and $\ell_{-i}(\cdot)$ is the log PL computed excluding sample $i$. $\lambda$ is chosen to maximize $\widehat{\text{CV}}_{\text{PL}}(\lambda)$. The CV log PL measures the difference of log PL on the full data and the log PL on the data excluding sample $i$, summed across all samples providing a goodness of fit estimate.
Note that  does not suffer from the sample size sensitivity outlined in the previous section, because it is used to compare models fit on the same sample size.

\section{Nested CV Algorithm}\label{sec3}
\begin{figure}
\centerline{\includegraphics[width=500pt,height=25pc]{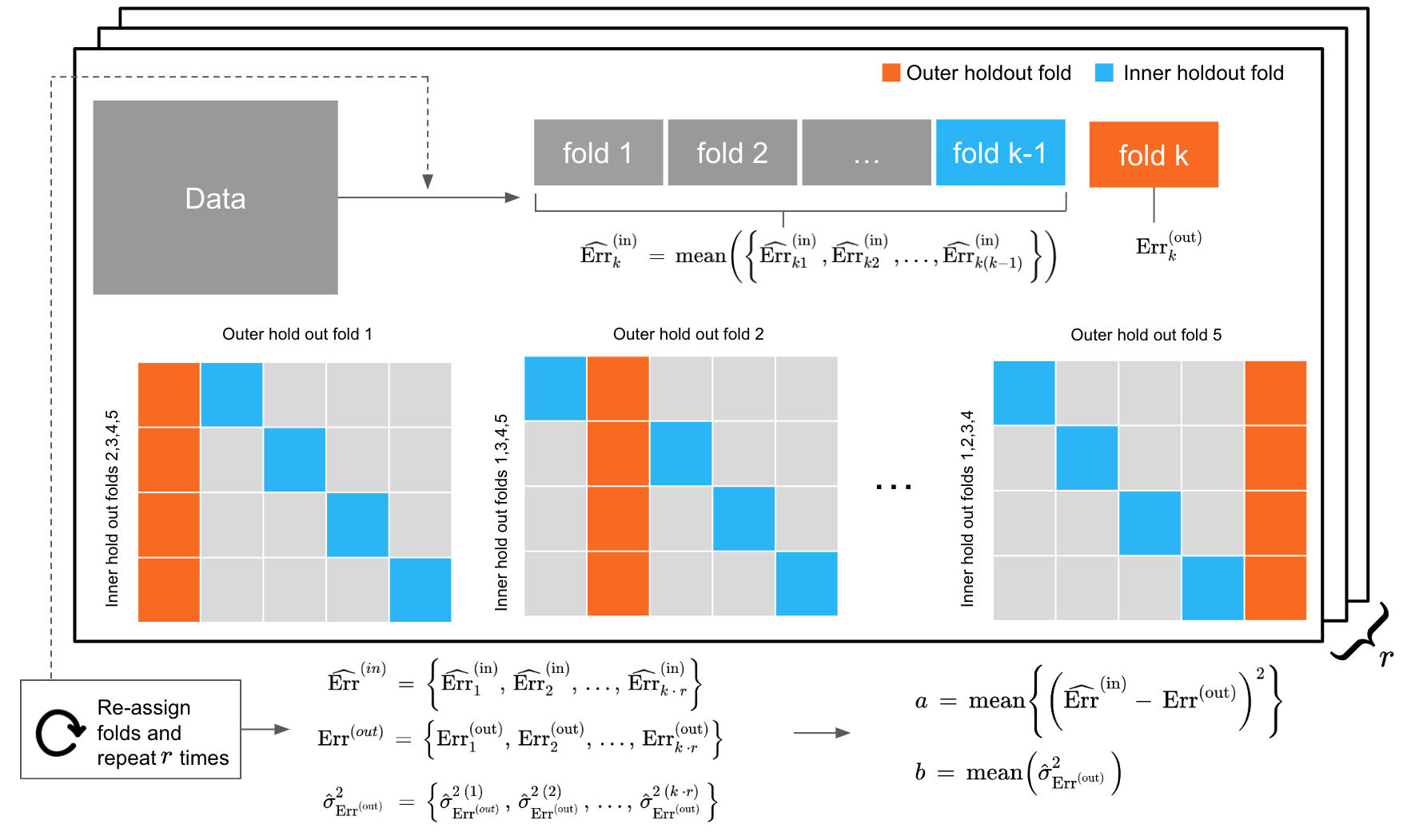}}
\caption{ {\it This diagram demonstrates how to compute the components $a$ and $b$ to estimate the MSE of the prediction error through the nested CV algorithm. To perform nested CV, the input data (typically the training data from random train-validation split) will be randomly split into $K$ equally sized folds. At each iteration of the procedure, one fold will be held out (represented in red) and run standard K-fold CV on the remaining $K-1$ folds (represented by the blue inner holdout fold and the grey folds). We repeat this procedure for all folds. This entire process is repeated $r$ times in order to get a reliable estimate of the MSE. Note that the 4x5 matrix is an example for when $K=5$.}  }
\end{figure}

Constructing confidence intervals for the prediction error estimate using naive CV leads to smaller intervals with poor coverage. Instead of using the empirical standard deviation of the CV estimates, we estimate the MSE of the prediction error using the MSE identity introduced by Bates {\it et al.} (2021), which allows the MSE to be directly estimated from given data as shown in Figure 3 \citep{Bates_2021}. The MSE is a good choice as it can be decomposed into bias and variance, where bias tends to be small, making it a close estimate of the variance. Bates {\it et al.} (2021) defines the MSE identity as follows,
\begin{eqnarray}
\label{eqn:MSE}
\underbrace{\mathbb{E}[(\widehat{\text{Err}}_{XY}^{(in)} - \text{Err}_{XY})^{2}]}_\text{MSE} = \underbrace{\mathbb{E}[(\widehat{\text{Err}}_{XY}^{(in)} - \bar{e}^{(out)})^{2}]}_\text{a} - \underbrace{\mathbb{E}[(\text{Err}_{XY} - \bar{e}^{(out)})^{2}]}_\text{b}
\end{eqnarray}
where $\widehat{\text{Err}}^{(in)}$ is the CV estimate of the errors from inner CV, $\bar{e}^{out}$ is the average error from the held out folds, and $\text{Err}_{XY}$ is the true prediction error conditioned on the training set $X$, $Y$. To compute part $a$ of the MSE estimate, we take the mean of $\widehat{\text{Err}}^{(in)}$ and $\bar{e}^{out}$ from multiple repetition then take the squared difference. For each iteration, the observations are randomly assigned to the $K$ folds. Part $b$ of the MSE is simply the variance of the heldout error $\bar{e}^{out}$. For the C-index, we estimate the variance using the infinitesimal jackknife estimate. The NCV point estimate $\widehat{\text{Err}}^{\text{(NCV)}}$ is the mean of all $\widehat{\text{Err}}^{\text{(in)}}$. The bias of $\widehat{\text{Err}}^{\text{(NCV)}}$ can also be estimated through the nested CV procedure,
\begin{eqnarray}
\widehat{\text{bias}} = \left(1 + \frac{K-2}{K} \right)\left[\widehat{\text{Err}}^{(\text{NCV})} - \widehat{\text{Err}}^{(\text{CV})} \right]
\end{eqnarray}
where the term $\widehat{\text{Err}}^{\text{(NCV)}} - \widehat{\text{Err}}^{\text{(CV)}}$ is the unbiased estimate of the difference in errors of sample size $n(K-2)/K$ and $n(K-1)/K$ respectively. The scaling factor $1 + (K-2)/K$ is included to account for biases going from different sample sizes. $1$ is for scaling back to an estimate of sample size $n(K-1)/K$ from $n(K-2/K)$ as the point error estimate from inner CV uses a sample size of $n(K-2/K)$. The term $(K-2)/K$ is for re-scaling the point estimate back to an estimate of size $n$ from $n(K-1/K)$.

Once we compute the point, MSE, and bias estimates through the nested CV procedure, we can construct the confidence interval using the following formula,

\begin{eqnarray}
\widehat{\text{Err}}^{\text{(NCV)}} - \widehat{\text{bias}} \pm q_{1-\frac{\alpha}{2}} \cdot \sqrt{\frac{K-1}{K}} \cdot \sqrt{\widehat{\text{MSE}}}
\end{eqnarray}
where $q_{1-\frac{\alpha}{2}}$ is the $1-\frac{\alpha}{2}$ quantile of the standard Gaussian. As $\widehat{\text{MSE}}$ is an estimate of a $K-1$ fold CV with $n(K-1)/K$ samples, the factor $\sqrt{K-1/K}$ is included to re-scale back to an estimate of sample size of $n$ , i.e. use $(K-1/K) \cdot \widehat{\text{MSE}}$.

\section{Experiments}\label{sec4}
\begin{figure}
\centerline{\includegraphics[width=500pt,height=14pc]{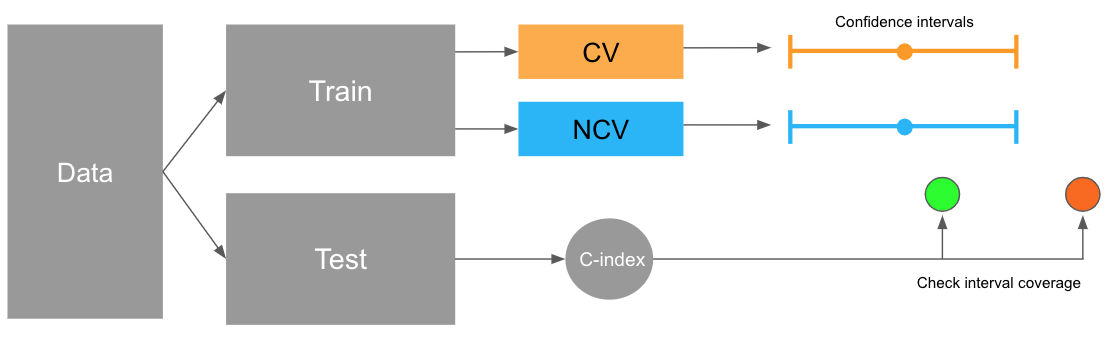}}
\caption{ {\it The simulation experiment process for one iteration. Data is split into train and test set randomly. We use the train set to run CV and NCV for constructing confidence intervals. We use the test set to compute the true test error (in this case C-index, but can also be PL or other measures). We repeat this procedure for a large number of times to compute the miscoverage rates of both CV and NCV based confidence intervals. Miscoverage rate is simply the count of the test set C-index falling within the CV or NCV confidence intervals. For all the experiments we aim for $90\%$ confidence intervals, which means that a valid $90\%$ confidence interval would have around $10\%$ miscoverage rate.}}
\end{figure}

We explore and demonstrate the efficacy of NCV in both numerical simulations and real survival data problems by running experiments to assess the coverage of confidence intervals constructed using the nested CV approach. Figure 4 delineates the experimental setup. The simulated or real data set is randomly split into train and test set. The train set is used to estimate both the standard CV confidence interval and the NCV confidence interval, while the test set is used to measure the test error. To evaluate whether the intervals achieve the desired $(1-\alpha) \%$ coverage, we count the number of times the test error is not contained by the interval and call this the interval's miscoverage rate. For instance, a miscoverage rate of $10\%$ is desired for a $90\%$ confidence interval. This would mean for a simulation of $100$ iterations, we should approximately expect to see $10$ of the intervals to not contain the test error. In our simulations, we primarily focus on the C-index, which is one of the most commonly employed metrics to evaluate the Cox model performance. 
    
Throughout all our experiments, our estimand of interest is equation \ref{eqn:MSE}, the mean squared error of the model prediction error, which is estimated from the training data through the nested CV procedure and used to construct the confidence intervals. We vary the number of samples $n$, number of features $p$, number of CV fold $n_{\text{fold}}$, and number of simulations $n_{\text{sim}}$. For larger $p > n$, we chose smaller values of $n$, $n_{\text{fold}}$, and $n_{\text{sim}}$ to reduce the computational cost resulting from relatively larger $p$.

The following experiments were implemented in {\bf R} V4.0.2 and {\bf glmnet} V4.1 for CV and fitting Cox models. Despite being more computationally expensive than CV, the nested CV algorithm can be implemented to compute the repetitions in parallel. These experiments were run on nodes with 64GB memory and 20-32 cores on Sherlock, a Stanford Research Computing Center (SRCC) HPC cluster. All code for the experiments are available on: \url{https://github.com/minwoosun/nestedcv-confint}.


\subsection{Simulation study}

We begin our simulation experiments with a simple base case with no censoring then extend to well-known parametric survival models and also incorporate right censoring. For each approach we cover both $n > p$ and $p > n$ settings. We describe in detail below our data-generating mechanisms for simulated data.

\begin{itemize}
    \item \textbf{Simple base case}

     We generate data $X \in \mathbb{R}^{n,p}$ with a signal such that time $ t =  x\beta + c \varepsilon $, where $ \varepsilon \stackrel{iid}{\sim} \mathcal{N}(0,1)$ and a constant $c \in \mathbb{R}$ which allows us to vary the degree of noise. The covariates are generated such that $p \stackrel{iid}{\sim} \mathcal{N}(0,1)$. We choose alpha to be $0.10$, i.e. construct $90\%$ confidence intervals. For the n > p scenario, we generate a train data set with n=100, p=10, and a large test data set with n=1000, p=10. For the p > n scenario, we generate a train data set of size n=100, p=150 and a large test data set with n=1000, p=150. We perform 10-fold CV and 10-fold nested CV fitting Cox PH model repeated 200 times for each simulation. We run a total of 1000 simulations for the n > p setting and 100 simulations for p > n setting.

    \item \textbf{Parametric models for survival time}

    To simulate more realistic survival times for the Cox proportional hazards model, we use the inverse transform sampling method, which is a general technique for generating random variables with a given cumulative distribution function. In particular, we follow the approach described in Bender et al. \citep{Bender_Augustin_Blettner_2005}. This method works in two stages: (1) sample from a uniform distribution (2) use the inverse of the cumulative distribution function of the target distribution to transform the uniform samples into the desired distribution like the Weibull or Gompertz distribution. Let $u$ be the sample from a uniform distribution, $u \sim \mathcal{U} [0 ,1]$, $x$ be the data matrix, and $\beta$ be the vector of values representing the strength of true association between the variables and the survival time. We generate the survival times for the exponential distribution with parameter $\lambda$ as,

    \begin{eqnarray}
        T_{\text{exponential}} = - \frac{\log(u)}{\lambda e^{\beta^{\top} x}} 
    \end{eqnarray}

    and for the Weibull distriubtion with parameters $\lambda$ and $\nu$ as,
    
    \begin{eqnarray}
        T_{weibull} = \left(- \frac{\log(u)}{\lambda e^{\beta^{\top} x}}\right)^{1/\nu} 
    \end{eqnarray}

    This allows for more flexibility in simulating survival times under different hazard functions and baseline hazard distributions. We also incorporate fixed censoring by setting a time ceiling, i.e. any time sampled past the max time will be considered right censored. 
    For the exponential survival time simulation, each continuous variable $X_{j}$ is sampled i.i.d. from a Gaussian distribution with mean $\mu=10$ and variance $\sigma^{2}=5$. We set $\beta = \{0.1, 0.1, 0, \cdots, 0\}$. For the Weibull survival time simulation, each binary variable $X_{j}$ is sampled i.i.d. as $n$ independent Bernoulli trials with probability of success $0.5$. For $n > p$ we let $n=1000$ and $p=10$. For $p > n$ we let $n=200$ and $p=250$. We set $\beta = \{0.1, 0.1, 0, \cdots, 0\}$ to induce moderate association between the features and the survival times. This set of simulation parameters leads to an average of $0.7$ CV C-index. 
\end{itemize}


Table 1 displays the findings for all the simulation experiments. We observe that the nested CV method attains valid coverage. In contrast, the basic CV strategy does not yield satisfactory results. In line with the regression and classification scenarios described in Bates {\it et al.} (2021), it becomes apparent that the miscoverage of the naive CV are significantly amplified when the feature count exceeds the observation count.

\subsection{Real data examples}
In addition to the simulated data, we compare the CV and nested CV procedures on real data sets. Similar to the simulated data setting, for each iteration we randomly sample a small number of observations from the entire data for the train set to run CV and nested CV for constructing confidence intervals. We use the remaining observations as the test set to evaluate the C-index and check whether the value falls within the confidence interval. Unlike in the simulation settings, we are no longer able to control the strength of association between the survival time and the features as the data generating mechanism is unknown. Through this repeated sampling procedure, we demonstrate the efficacy of the NCV procedure beyond the simulation setting in two real data sets representing both the $n > p$ and $p > n$ scenarios. To be comprehensive, we choose a clinical trial data set and a cancer gene expression dataset, which are both commonly found data modalities used to address modern biomedical problems.

\begin{itemize}
    \item \textbf{Chemotherapy for stage III colon cancer data (n > p)}
    
    The colon cancer data comes from a randomized control trial that aimed to assess the efficacy of combining two adjuvant therapy for stage III colon cancer \citep{Moertel_1995}. The data set comprises 929 patients and 11 predictors. The data matrix contains a total of 1858 rows as there are two observations per patient for different event types: death and recurrence. The predictors include the treatment type, clinical information like the patient age and sex, and quantitative descriptions of the tumor and colon. We filter for event type death (etype=2) leading to a data matrix with n=929 and p=11. We ran a total of 200 simulations with 200 repetitions.  

    \item \textbf{Prediction of Clinical Outcomes from Genomic Profiles (PRECOG) data (p > n)}

    The PRECOG data set is an amalgamation of 165 cancer gene expression data with patient clinical outcome \citep{Gentles_2015}. This pan-cancer data set was curated with the goal of better understanding the prognostic landscape of genes and immune cells for humans, which can lead to novel biomarker discoveries and therapeutic targets. Gene expression data are typically characterized by a large number of genes to be used as features with a much smaller number of samples. For the experiment, we chose and preprocessed the 8 data sets that were used to train the FOXM1-KLRB1 prognostic model as described in the paper resulting in n=1086, p=4708. The ID for gene expression data sets used for training are GSE32062, E-TABM-38, GSE10358, GSE4475, GSE8894, GSE1993, and GSE19234. We converted the expression values into z-scores then took the top 180 features with the highest variance to reduce the feature space leading to the final matrix of size n=1086, p=180. We randomly sampled 150 patients from the 1086 patients for constructing the confidence intervals and used the remainder for measuring the C-index. We ran a total of 150 simulations with 200 repetitions each. 
    
\end{itemize}

The results for the real data experiments are reported in Table 2. In both the colon cancer chemotherapy data set $n > p$ and the PRECOG data set  $p > n$, the nested CV confidence intervals achieve proper coverage, while the naive CV approach fails. Similar to the simulation experiments, the naive CV miscoverage is much worse when there are more features than the number of observations. These findings further validate how effective the nested CV procedure is for constructing valid confidence intervals.

\section{Discussion}\label{sec5}
In this study, we set out to underscore the importance and efficacy of the nested CV procedure as a reliable approach for estimating the MSE of the prediction error and constructing valid confidence intervals for the Cox model test error. Our aim was to show that this method leads to confidence intervals of the prediction error that achieve desired coverage, which the naive CV approach may fail to attain. 

To validate our theory, we conducted a series of simulations and experiments with real-world data. The purpose of these experiments was twofold: First, to substantiate our claim about the nested CV procedure's efficacy in attaining valid coverage, and second, to demonstrate the limitations of naive CV based confidence intervals. The results showed that naive CV based confidence intervals indeed had poor coverage rates due to the interval width being too narrow. This shortcoming was even more pronounced in the $p > n$ scenario, where the number of features p far exceeds the number of observations n consistently throughout all the experiments. In contrast to the naive CV confidence intervals, the results of the nested CV procedure were far more satisfactory, and indeed aligned with our expectations. The confidence intervals derived from the nested CV procedure were shown to achieve the nominal coverage, a key indication of the method's reliability. This is a significant finding because it points to the robustness of the nested CV procedure in the face of increasing complexity and high-dimensional data.

While we focused on the C-index for the experiments, this approach is not limited to the C-index metric alone. Bates {\it et al.} (2021) has shown that nested CV can be used for model performance metrics like the MSE in regression settings and accuracy in classification settings. Furthermore, the nested CV algorithm can be applied to other models beyond just the Cox PH model, including other supervised learning models through survival stacking \citep{craig2021}. However, the exact conditions for when the standard CV intervals fail even in the $n > p$ setting is still not fully understood. Generally, we observed that increasing the sample size $n$ led to better coverage. We also saw that a sufficiently large number of repetition for the nested CV algorithm is necessary for getting a reliable estimate of the MSE for $p >n$ scenarios and for the confidence intervals to attain proper coverage. Increasing the number of repetition renders the process computationally more intensive, but these repetitions can be easily run in parallel, mitigating the computational challenges. This is an important advantage, as it allows for the process to be still viable for large-scale applications.

In conclusion, our work attests to the effectiveness of the nested CV procedure for estimating the MSE of the prediction error and for constructing confidence intervals of the Cox model test error. This method outperforms the standard CV approach, which our experiments revealed as less reliable, especially when the number of features exceeds the number of observations.

\section*{Acknowledgements}\label{sec6}
We thank Dr. Lu Tian for his contribution on deriving the expectation of $\frac{\log(PL)}{n}$. M.S. was supported by the National Library of Medicine (LM007033);
 R.T. was supported by the National Institutes of Health (5R01 EB001988-16)
and the National Science Foundation (19 DMS1208164).

\begin{center}
\begin{table}[]
\centering
\caption{{\it This table summarizes the results from all the simulated data examples. We report the point estimate, mean standard error across multiple iterations and the confidence interval miscoverage rates.}}
\begin{tabular}{|l|cc|cc|clclclcl|}
\hline
\multicolumn{1}{|c|}{\textbf{Setting}} & \multicolumn{2}{c|}{\textbf{Point Estimate}} & \multicolumn{2}{c|}{\textbf{Mean SE}} & \multicolumn{8}{c|}{\textbf{Miscoverage}}                                                                       \\
\multicolumn{1}{|c|}{}                 &                                 &            &                              &        & \multicolumn{4}{c}{CV}                                 & \multicolumn{4}{c|}{NCV}                               \\
\multicolumn{1}{|c|}{}                 & CV                              & NCV        & CV                           & NCV    & \multicolumn{2}{c}{Upper} & \multicolumn{2}{c}{Lower}  & \multicolumn{2}{c}{Upper} & \multicolumn{2}{c|}{Lower} \\ \hline
Base case (n \textgreater p)           & \multicolumn{1}{c|}{0.605}      & 0.594      & \multicolumn{1}{c|}{0.037}   & 0.053  & \multicolumn{2}{c}{0.190} & \multicolumn{2}{c|}{0.030} & \multicolumn{2}{c}{0.060} & \multicolumn{2}{c|}{0.010} \\ \hline
Base case (p \textgreater n)           & \multicolumn{1}{c|}{0.667}      & 0.661      & \multicolumn{1}{c|}{0.027}   & 0.063  & \multicolumn{2}{c}{0.350} & \multicolumn{2}{c|}{0.013} & \multicolumn{2}{c}{0.050} & \multicolumn{2}{c|}{0.080} \\ \hline
Exponential (n \textgreater p)         & \multicolumn{1}{c|}{0.676}      & 0.676      & \multicolumn{1}{c|}{0.009}   & 0.014  & \multicolumn{2}{c}{0.050} & \multicolumn{2}{c|}{0.090} & \multicolumn{2}{c}{0.040} & \multicolumn{2}{c|}{0.050} \\ \hline
Exponential (p \textgreater n)         & \multicolumn{1}{c|}{0.673}      & 0.654      & \multicolumn{1}{c|}{0.019}   & 0.037  & \multicolumn{2}{c}{0.060} & \multicolumn{2}{c|}{0.240} & \multicolumn{2}{c}{0.040} & \multicolumn{2}{c|}{0.020} \\ \hline
Weibull (n \textgreater p)             & \multicolumn{1}{c|}{0.675}      & 0.675      & \multicolumn{1}{c|}{0.006}   & 0.008  & \multicolumn{2}{c}{0.070} & \multicolumn{2}{c|}{0.130} & \multicolumn{2}{c}{0.070} & \multicolumn{2}{c|}{0.040} \\ \hline
Weibull (p \textgreater n)             & \multicolumn{1}{c|}{0.674}          & 0.656          & \multicolumn{1}{c|}{0.019}       & 0.033      & \multicolumn{2}{c}{0.080}     & \multicolumn{2}{c|}{0.160}     & \multicolumn{2}{c}{0.080}     & \multicolumn{2}{c|}{0.020}     \\ \hline
\end{tabular}
\end{table}
\end{center}

\begin{center}
\begin{table}[]
\centering
\caption{{\it This table summarizes the results from all the real data examples. We report the point estimate, mean standard error across multiple iterations and the confidence interval miscoverage rates.}}
\begin{tabular}{|l|cc|cc|clclclcl|}
\hline
\multicolumn{1}{|c|}{\textbf{Setting}} & \multicolumn{2}{c|}{\textbf{Point Estimate}} & \multicolumn{2}{c|}{\textbf{Mean SE}} & \multicolumn{8}{c|}{\textbf{Miscoverage}}                                                                       \\
\multicolumn{1}{|c|}{}                 &                                 &            &                              &        & \multicolumn{4}{c}{CV}                                 & \multicolumn{4}{c|}{NCV}                               \\
\multicolumn{1}{|c|}{}                 & CV                              & NCV        & CV                           & NCV    & \multicolumn{2}{c}{Upper} & \multicolumn{2}{c}{Lower}  & \multicolumn{2}{c}{Upper} & \multicolumn{2}{c|}{Lower} \\ \hline
Colon (n \textgreater p)               & \multicolumn{1}{c|}{0.656}      & 0.651      & \multicolumn{1}{c|}{0.026}   & 0.036  & \multicolumn{2}{c}{0.145} & \multicolumn{2}{c|}{0.025} & \multicolumn{2}{c}{0.085} & \multicolumn{2}{c|}{0.015} \\ \hline
PRECOG (p \textgreater n)              & \multicolumn{1}{c|}{0.697}      & 0.671      & \multicolumn{1}{c|}{0.035}   & 0.065  & \multicolumn{2}{c}{0.173} & \multicolumn{2}{c|}{0.040} & \multicolumn{2}{c}{0.020} & \multicolumn{2}{c|}{0.087} \\ \hline
\end{tabular}
\end{table}
\end{center}

\bibliographystyle{agsm} 
\bibliography{references}

\end{document}